\documentclass{ieeeaccess}

\usepackage{verbatim}

\usepackage{cite}
\usepackage{amsmath,amssymb,amsfonts}
\usepackage{algorithmic}
\usepackage{graphicx}
\usepackage{subfig}
\usepackage{textcomp}

\def\BibTeX{{\rm B\kern-.05em{\sc i\kern-.025em b}\kern-.08em
    T\kern-.1667em\lower.7ex\hbox{E}\kern-.125emX}}

\begin{document}
\history{Date of publication xxxx 00, 0000, date of current version xxxx 00, 0000.}
\doi{10.1109/TQE.2020.DOI}

\title{Hybrid classical--quantum branch--and--bound algorithm for solving integer linear problems}
\author{\uppercase{Claudio Sanavio}\authorrefmark{1,2,3}, 
\uppercase{Edoardo Tignone\authorrefmark{4}, and Elisa Ercolessi}\authorrefmark{2,3}
}
\address[1]{Fondazione Istituto Italiano di Tecnologia, 
Center for Life Nano-Neuroscience at la Sapienza, 
Viale Regina Elena 291, 00161 Roma, Italy}
\address[2]{Dipartimento di Fisica e Astronomia dell'Universit\`a di Bologna, I-40127 Bologna, Italy}
\address[3]{INFN, Sezione di Bologna, I-40127 Bologna, Italy}
\address[4]{Leithà S.r.l.~\text{\textbar} Unipol Group, Bologna, Italy}
\tfootnote{This research is funded by the International Foundation Big Data and Artificial Intelligence for Human Development (IFAB) through the project “Quantum Computing for Applications”.
E.~E.~is partially supported by INFN through the project “QUANTUM”. C.~S. and E.~E. acknowledge financial support from the National Centre for HPC, Big Data and Quantum Computing (Spoke 10, CN00000013).}

\markboth
{Claudio Sanavio \headeretal: Hybrid classical--quantum branch--and--bound algorithm ...}
{Claudio Sanavio \headeretal: Hybrid classical--quantum branch--and--bound algorithm ...}

\corresp{Corresponding author: Claudio Sanavio (email: claudio.sanavio@iit.it).}

\begin{abstract}
Quantum annealers are suited to solve several logistic optimization problems expressed in the QUBO formulation. However, the solutions proposed by the quantum annealers are generally not optimal, as thermal noise and other disturbing effects arise when the number of qubits involved in the calculation is too large. In order to deal with this issue, we propose the use of the classical branch--and--bound algorithm, that divides the problem into sub-problems which are described by a lower number of qubits. We analyze the performance of this method on two problems, the knapsack problem and the traveling salesman problem. Our results show the advantages of this method, that balances the number of steps that the algorithm has to make with the amount of error in the solution found by the quantum hardware that the user is willing to risk. All the results are actual runs on the quantum annealer D-Wave Advantage.
\end{abstract}


\titlepgskip=-15pt

\maketitle

\section{Introduction}
\label{sec:introduction}

Logistic optimization problems, whose goal is to find the solution which minimizes a suitable cost function given a set of constraints, can often be expressed in terms of binary combinatorial problems. 
One way to tackle this kind of problems consists of exploring all possible solutions, thus pursuing a brute-force strategy. A better strategy is found in the so called branch--and--bound (BB) algorithm~\cite{Land1960}, that explores sub-combinations of the problem and excludes those that either do not satisfy the constraints or those whose value of the cost function is higher than the solutions previously investigated. Paradigmatic examples, that can be both solved with the BB algorithm~\cite{Kolesar1967,Laporte1983,Toth2002,MartelloToth1990}, are given by: \\
i) the Knapsack Problem (KP)~\cite{MartelloToth1990}, in which one searches for the selection of objects (from a predefined set) that maximizes the load's value while adhering to the capacity constraint of the carrier;\\
ii) the Traveling Salesman Problem (TSP)~\cite{Toth2002,Laporte1992}, which aims at finding the minimal route that passes through different cities, with the constraint that the path crosses all of the cities exactly once. 

In recent times, with physical platforms that have been made available to researchers, attention has been driven by the possibility that quantum computers can speed-up the resolution of several NP-hard problems~\cite{NielsenChuang2000}. In this work we analyze a particular quantum computer, the quantum annealer of D-Wave, which is expected to be particularly suitable for solving binary optimization problems. The two problems mentioned above are perfect examples of combinatorial problems that may benefit from the use of a quantum computer. However, few works have been done on the actual resolution on the D-wave machine.  In Ref.~\cite{Pusey2020} the authors tried to solve small KP using the D-wave quantum annealer, finding results far from the global optimum. In Ref.~\cite{Krauss2020} the authors have analyzed the shortest path problem, a slight variant of the TSP. They successfully determined the optimal solution for graphs composed of up to six nodes, a limitation that may not align with current practical requirements. In fact, the quantum annealer is still a developing technology, currently affected by noise that makes it unable to find the global solution even to simple problems.

In order to overcome the apparent failure of the fruitful use of this device in the near period, we propose to use a classical-hybrid protocol, in which the quantum hardware is used as a subroutine for a variant of the classical BB algorithm. 
In particular, we show that thanks to the BB algorithm we can reduce the size of the problem down to a number of instances that are feasible for the quantum computer. With this strategy, the quantum computer can fully show its potential and find an optimal solution to the problem. \\
Our findings show that we can exploit near-term quantum computers to speed up the solution of a problem reducing the number of queries to both the quantum and the classical computer. However, there is a trade off between the quality of the found solution, measured in terms of its proximity to the global optimal solution, and the achievable speed-up. A similar idea has been proposed in Ref.~\cite{Rosenberg2016}, the BB algorithm was applied to a large number of QUBO problems, by simulating the quantum annealer on a classical computer. 

In the paper, we apply a hybrid algorithm to study both the KP and the TSP making use of the real D-Wave machine Advantage~\cite{Dwave1,Dwave2,Dwave3}. The content is as follows.
In Section~\ref{sec:II} we review the definition of  the TSP and the KP and we explain how they can be solved using the BB algorithm. In Section~\ref{sec:III} we show how we can encode the optimal solution of the TSP and the KP as the ground state of a suitable Hamiltonian for the quantum annealer. In Section~\ref{sec:IV} we first discuss the issues related to directly solve the integer linear problems via unrestricted algorithm on the quantum hardware. Next,  we define our hybrid classical-quantum algorithm and study its performance on examples of the KP and TSP problems. In Section~\ref{sec:V} we draw our conclusions and present outlooks.

\section{The binary linear problem}\label{sec:II}

The binary linear optimization problem (BLOP) aims to find the minimum of the cost function $z(\mathbf{x})$ over the set of possible solutions $\Omega$, namely

\begin{subequations}\label{eq:BLproblem}
\begin{eqnarray}\label{eq:BLproblema}
& \min_{\mathbf{x}\in\Omega}\; (\mathbf{x})=\mathbf{c}^T\mathbf{x}\\
\text{s.t.}&A\mathbf{x}\leq \mathbf{b}\label{eq:BLproblemb}\\
&\mathbf{x}\in\{0,1\}^{N},\label{eq:BLproblemc}
\end{eqnarray} 
\end{subequations}

\noindent with $\mathbf{c}$ being an $N$-dimensional vector, $A$ being an $N\times m$ matrix and $\mathbf{b}$ being an $m$ dimensional vector. $\mathbf{x}$ is an $N$ dimensional vector whose components take value $0$ or $1$. Although the simple form of Eqs.~\eqref{eq:BLproblem}, the BLOP is generally NP-hard~\cite{Papadimitriou1981CombinatorialOA}.

In this section we explore one algorithm that is used to solve BLOPs, the BB algorithm. 
The core procedure of the BB algorithm for finding the solution of the problem $\mathcal{P}_{\Omega}$ on the set $\Omega$ consists into analyzing the restrictions $\mathcal{R}_{\bar{\Omega}}$ of the original set $\bar{\Omega}\subset\Omega$. 

\noindent The lowest value of the cost function of the original problem $\min_{\mathbf{x}\in\Omega} z(\mathbf{x})$ is a lower bound of the minimum of the cost function of the restricted problem $\min_{\mathbf{x}\in\bar{\Omega}} z(\mathbf{x})$. 
As we are dealing with a linear problem, the minimum and the maximum of the cost function is to be found at the borders of the set $\Omega$. 
However the optimal solution could have non-binary values. 

\noindent Dividing the original problem into sub-problems allows us to consider the restricted problems separately and choose the one with minimum cost.
Fig.~\ref{fig:BBa} shows the procedure of the BB algorithm when applied to a 3-dimensional system. In Fig.~\ref{fig:BBb} the procedure is represented as a tree, where each branch represents a restriction of the problem into a selected subset.
 
We define $z_{\mathcal{P}}$ as the \textit{current} upper bound of the cost function. We initialize it to the upper bound of the cost function over the set $\Omega$. This is done at the root of the tree and is equivalent to initialize $z_{\mathcal{P}}$ to the value $\infty$. 
In the example shown in Fig.~\ref{fig:BB}, we analyze the restricted problem $\mathcal{R}_{0}$, where $\bar{\Omega}$ is a lower dimensional set that has the value of the first variable fixed to zero, i.e. $x_1=0$. The lower bound of the cost function in the restricted region is $z_{\mathcal{R}_{0}}<z_{\mathcal{P}}$, because of the original large value of $z_{\mathcal{P}}$. 
We then proceed by restricting the second variable $x_2=0$ so defining the problem $\mathcal{R}_{00}$, and finally the third one $x_3=0$ in order to specify the problem $\mathcal{R}_{000}$, and find the value of the cost function $z(0,0,0)$ for this restriction.
This procedure is called branching and stands for the subsequent restriction of the problem into different sub-problems. Once the value of the cost function for a certain branch has been found, we proceed with the bounding procedure. 
The current upper bound $z_{\mathcal{P}}$ is updated to $z_{\mathcal{R}_{000}}$. From now on, we consider only subsets where the optimal cost function is less than the updated $z_{\mathcal{P}}$.
We now relax the problem, going back to $\mathcal{R}_{00}$, and set $x_3=1$. If $z_{\mathcal{R}_{001}}<z_{\mathcal{P}}$, the latter is updated to this value.

We analyze other values of the variables relaxing the problem and applying different restrictions, as long as we explore all the solutions. 
If the lower bound of the cost function $z_{\mathcal{R}_{i}}$ in the follow-up restriction $\mathcal{R}_{i}$ is greater than the current upper bound $z_{\mathcal{P}}$, we do not need to investigate that restricted region, and we can skip to another branch. 
If a node in the tree (a restricted subset) does not satisfy the constraints, the node is not considered valid and the bounding process does not take place.

\noindent In the tree of Fig.~\ref{fig:BBa} each branch develops on the assumption that the value of a variable is fixed to either $0$ or $1$, but the same method can be applied to integer values of the variables. 

\noindent Finally, the optimal solution will minimize the cost function while satisfying the constraints.

\begingroup
\begin{figure}
\centering
\subfloat[]{\includegraphics[width=.8\linewidth]{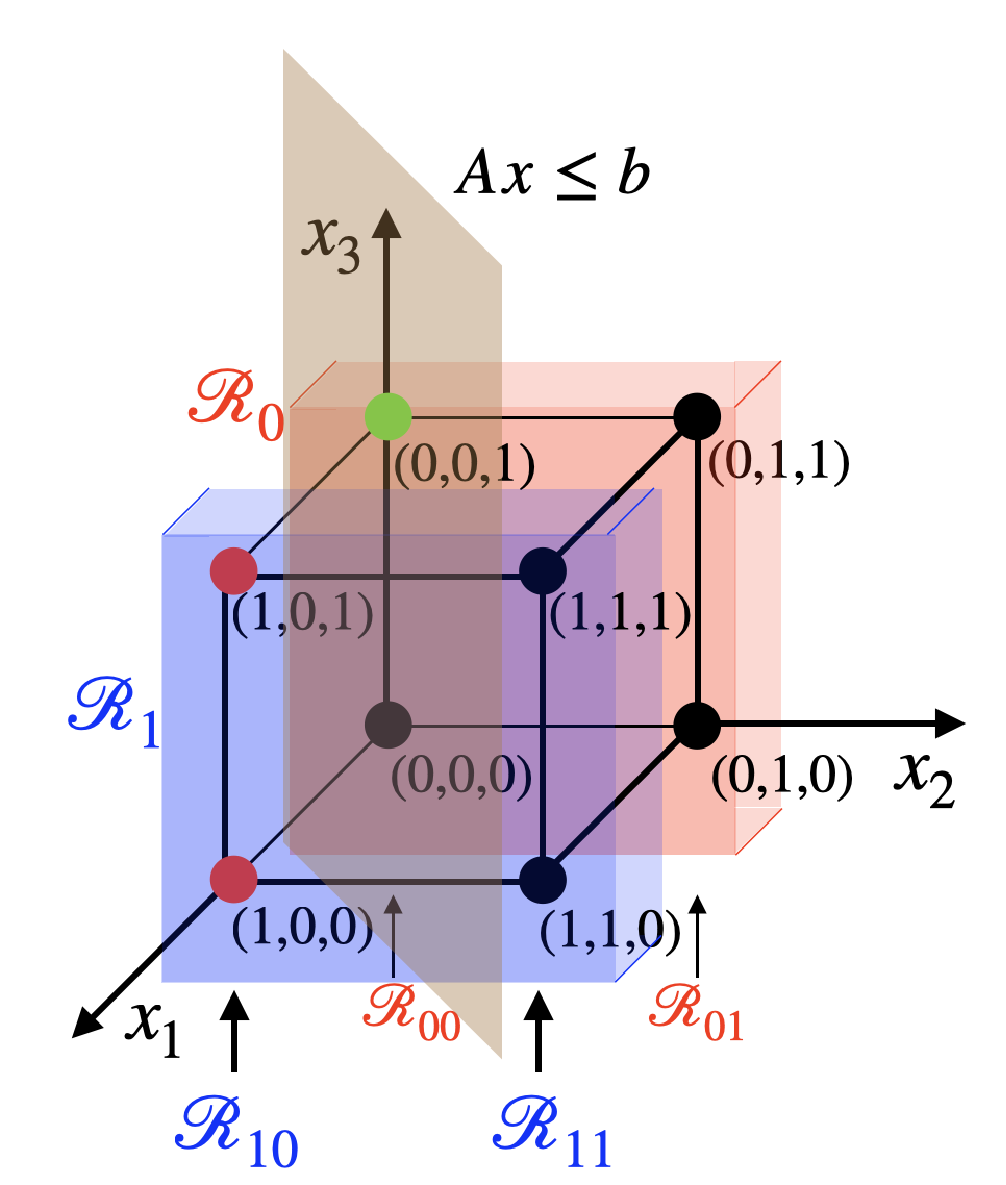}\label{fig:BBa}}

\subfloat[]{\includegraphics[width=\linewidth]{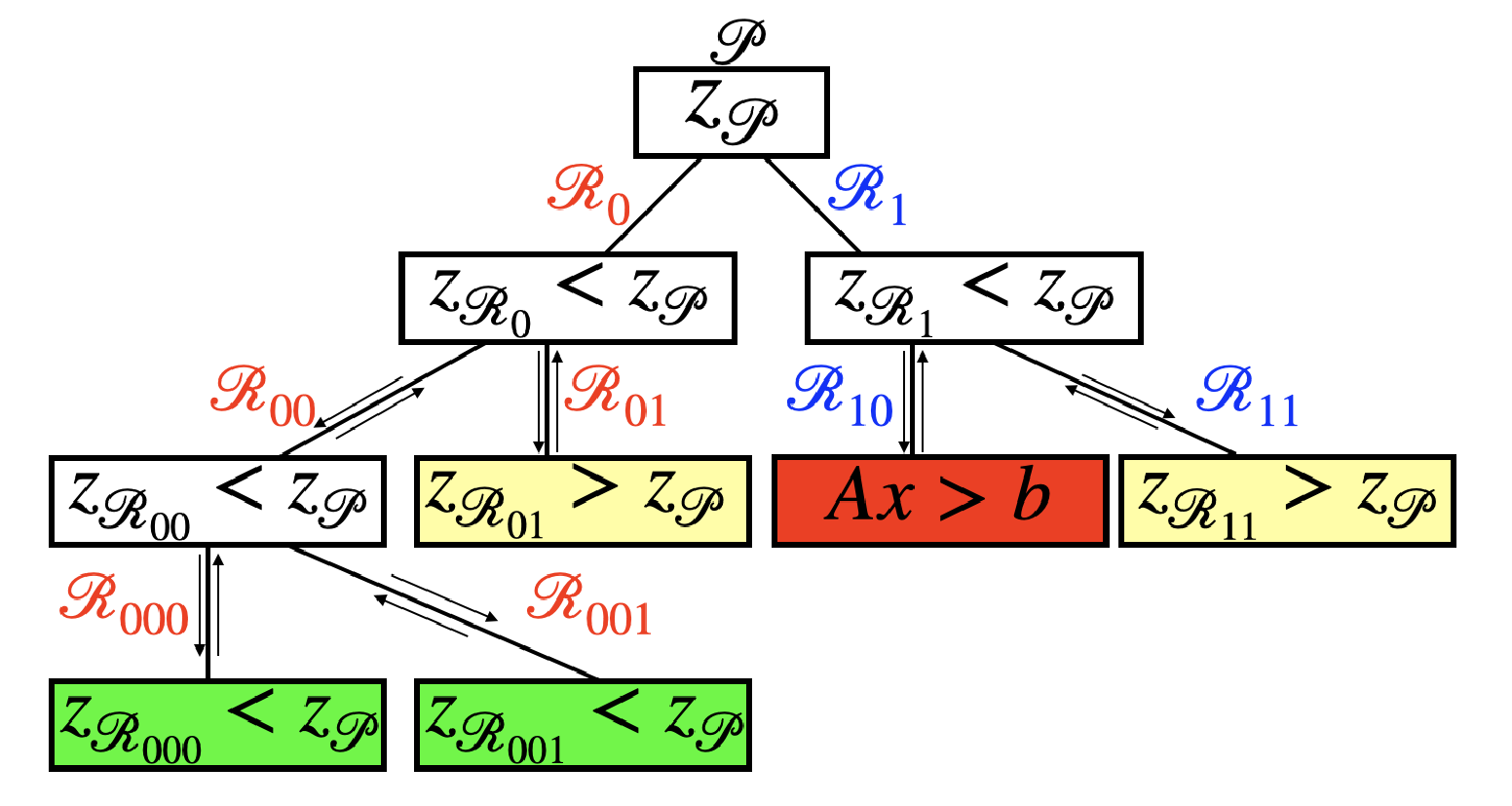}\label{fig:BBb}}

\caption{(a) Visualization of the BB algorithm in the 3 dimensional space $[0,1]^3$ with constraint $Ax\leq b$. Each restriction $\mathcal{R}$ represents a subset of $[0,1]^3$ where one or more variables have been constrained to a binary value. (b) Schematic representation of the BB algorithm as a tree. At each step BB checks if the constraints are satisfied and if the cost function of the node is lower than the current upper bound. When the branching has terminated, the bounding process takes place and the value of the upper bound $z_\mathcal{P}$ is updated (green node). If the next branch finds a value of $z>z_\mathcal{P}$ (yellow nodes) the branch is not explored further. The same happens if the constraints are violated (red node). \label{fig:BB}}
\end{figure}
\endgroup

\subsection{The knapsack problem}\label{sec:IIb}

In the KP, we have $N$ objects, each with a value $v_i$ and a weight $w_i$, for $i=1,\dots,N$. The problem consists of choosing the items to put in the knapsack so that their total value is maximum while not exceeding the knapsack capacity. This problem can be formulated by using $N$ binary variables $x_i=1$ ($i=1,\cdots,N$), with $x_i=1$ if the object $i$ enters in the knapsack, and $x_i=0$ if it does not. The values and the weights can be collected in two vectors $\mathbf{v}$ and $\mathbf{w}$ respectively. The BLOP \eqref{eq:BLproblem} is written as

\begin{subequations}\label{eq:KP}
\begin{eqnarray}
& \min_{\mathbf{x}} \; z(\mathbf{x})=-\mathbf{v}^T\mathbf{x}\label{eq:KPa}\\
\text{s.t.}&w(\mathbf{x}) = \mathbf{w}^T\mathbf{x}\leq W,\label{eq:KPb}\\
&\mathbf{x}\in\{0,1\}^N\label{eq:KPc}.
\end{eqnarray}
\end{subequations}

The KP can be solved using the BB algorithm. As we assume that all the weights $w_i$ and the values $v_i$ are positive numbers for all $i=1,\dots,N$, we can slightly modify the BB algorithm in order to make it more efficient. We call this version KP-BB. We start from the string $\mathbf{x}^{0}=(0,\dots,0)$ , which describes the empty knapsack. The value function $z$ and the weight function $w$ are both zero. The first branch has the value of the first variable set, $x_1=1$. The value function $z$ is updated to $-v_1$ and the weight function takes value $w_1$. The $k$-th branch sets the $k$-th variable $x_k=1$ and sets the previous $(k-1)$ variables to zero. At each $k$-th branch, the value function is $-v_k$ and the weight function is $w_k$. 
With this ordering, each of the $k$ branches defines a new knapsack problem, where the $k$-th object has been chosen and the problem is to choose among the remaining $N-k$ objects. The loading capacity of the new problem is $W-w_k$ and the value of the empty knapsack is initialized to $-v_k$. The KP-BB algorithm recursively applies this restriction of the problem to each branch, following a so-called depth-first search, where the search scouts a tree till it ends, then traces back its steps. 
At this point, the algorithm passes to another branch. 
In Fig.~\ref{fig:KPBB} we show the tree representation of the KP-BB algorithm for a problem with three objects. In the node's box the value of the upper bound of the cost function is shown. When a new solution is explored, the algorithm identifies it as optimal (green), valid but not optimal (yellow) or not valid (red).
It stops either when all the variables have been considered, or when the total weight of the knapsack has exceeded its limit. 
\begingroup
\begin{figure}[ht!]
\centering
\includegraphics[width=.8\linewidth]{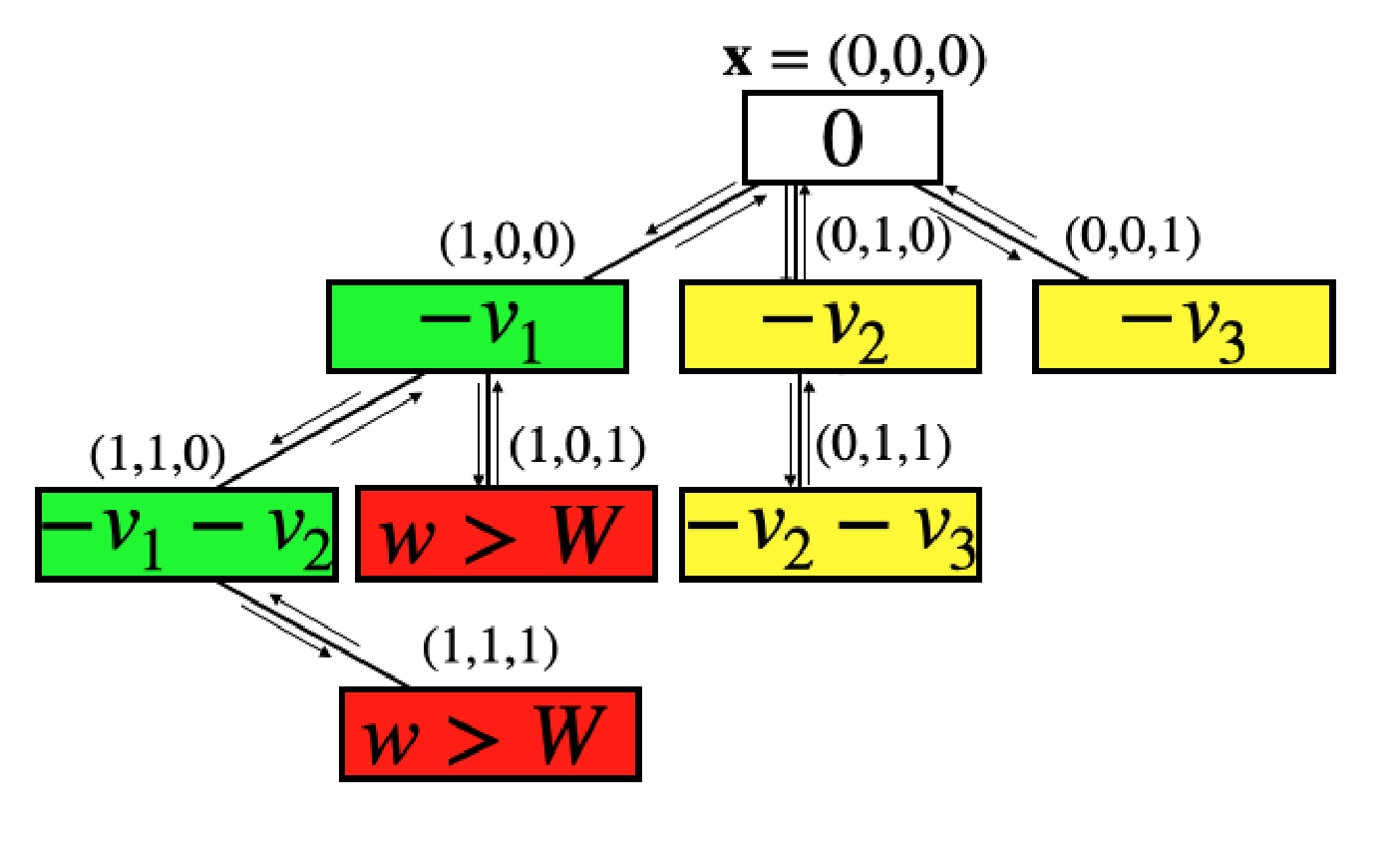}
\caption{Schematic representation of the KP-BB algorithm as a tree, when applied to a knapsack problem with three objects. The nodes represent a solution as written above them. In the boxes we write the value of the cost function for that solution. The knapsack is initially empty $\mathbf{x}=(0,0,0), z=0$ and it is filled with the first object, described by the solution vector $(1,0,0)$. The value function is updated to $-v_1$ and the weight to $w_1$. Hence, a new KP problem is defined as described in the text. A green node represents a valid solution that is currently optimal. A yellow node represents a solution that is valid, but not optimal. A red node represents a non-valid solution, where the constraints are not satisfied.\label{fig:KPBB}}
\end{figure}
\endgroup

\subsection{The traveling salesman problem}\label{sec:IIa}

The traveling salesman problem describes an agent that has to visit different cities starting from the depot and returning to it at the end of the journey. The problem consists of finding the route with the lowest cost that visits all the nodes exactly once, where the cost can describe time, distance or money used in the travel.  TSP can be modeled by a weighted graph, where the nodes are the cities and the weights of the edges represent the transportation cost from one node to another one. One can always assume the graph to be fully connected, by putting the corresponding cost to be very high (possibly infinity) if two cities are not actually connected by a direct path.

The graph is represented by the pair $G=(V,C)$ with $V=\{d,1,2,\dots,N-1\}$ being the set of $N$ vertices and $C$ being the weighted adjacency $N\times N$ matrix of the graph. The component $C_{ij}$ represents the cost of traveling from node $i$ to node $j$ and it is not necessarily symmetric. A solution for the TSP is the cycle path along the edges of the graph that starts from the depot node, named $d$, passes through all the other $N-1$ nodes exactly once and ends in the initial node $d$.
One mathematical description of TSP is given by the Dantzig--Fulkerson--Johnson formulation~\cite{DFJ1954}, which considers $N^2$ binary variables $x_{ij}$, each representing the edge that connects the node $i$ with the node $j$. The corresponding linear problem is obtained in the form of Eq.~\eqref{eq:BLproblem} collecting the binary variables $x_{ij}$ into a $N\times N$ matrix $\mathbf{x}$.

\begin{subequations}\label{eq:TSP}
\begin{eqnarray}
&\min_{\mathbf{x}}\; z(\mathbf{x})=\text{Tr}[C\mathbf{x}] = \sum_{i\neq j}C_{ij}x_{ij},\label{eq:TSPa}\\
\text{s.t.}&\sum_{j=1}^nx_{ij}=1,\quad \forall i\label{eq:TSPb}\\
&\sum_{i=1}^nx_{ij}=1,\quad \forall j\label{eq:TSPc}\\
&\sum_{i,j\in S}x_{ij}\leq \lvert S\rvert,\label{eq:TSPd}\\
&\forall S\subset V,\quad 2\leq\lvert S\rvert\leq N-2,\nonumber\\
&x_{ij}\in\{0,1\}\label{eq:TSPe}.
\end{eqnarray} 
\end{subequations}

The first line, Eq.~\eqref{eq:TSPa} is the objective function that we want to minimize. The other equations are the constraints given by the TSP. In particular, Eqs.~\eqref{eq:TSPb},\eqref{eq:TSPc} state that each vertex must have one inward and one outward edge respectively. Eq.~\eqref{eq:TSPd} avoids the presence of sub-paths that do not cover the whole set of vertices. Here $S$ is a subset of $V$, and $|S|$ is the number of elements in $S$. Eq.\eqref{eq:TSPe} states that the $x_{ij}$ are binary variables.

The TSP can be solved with the BB algorithm~\cite{Laporte1983,Toth2002}. 
Similar to the knapsack problem, the BB algorithm offers a strategy to systematically explore all solutions, which can be tailored for the specific problem at hand. We refer to this customized version of the branch--and--bound algorithm as TSP-BB.

\noindent We start from the depot with initial cost function $z=0$, and we choose the path to one of the $N-1$ cities, that we denote with $\overline{k} $ . Then, the TSP-BB algorithm defines a new TSP made of $N-2$ cities, where the cost matrix is modified to include the previous choice. The new TSP has $N-1$ vertices and new adjacency matrix $C'$ with adjacency components $C'_{i\overline{k}}=C_{id}$. Thus the initial cost of the new TSP has the updated value $z\to z+C_{d\overline{k}}$.
Different from the KP-BB defined before, the TSP-BB has a best-first search approach, where the next node analyzed by the algorithm is the one with current best $z$ value. When all the variables have been explored in one branch, we define the upper bound of the travel cost $z$. Any time the initial travel cost of a branch exceeds $z$, we neglect that branch.
On the contrary, we explore the branches with travel cost lower than $z$, until we find the optimal solution. 
Fig.~\ref{fig:TSPBB} shows a schematic representation of the TSP-BB algorithm as a tree, when applied to a graph with $5$ cities.

\begingroup
\begin{figure}[ht!]
\centering
\includegraphics[width=.8\linewidth]{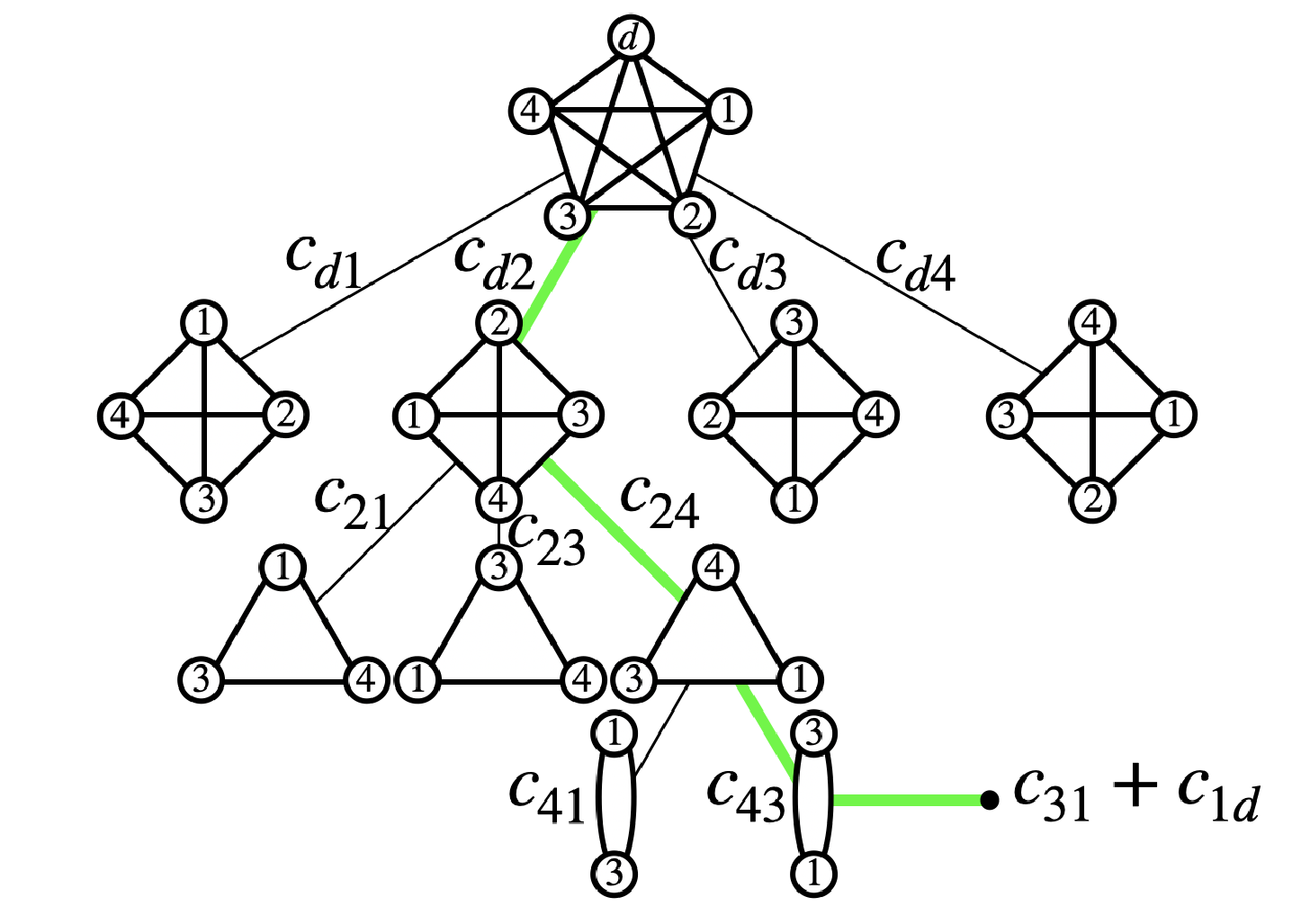}
\caption{Schematic representation of the TSP-BB algorithm applied to a traveling salesman problem with $5$ nodes. Once an edge is chosen, the problem reduces to another TSP with one node less. The initial travel cost is obtained summing up the partial travel costs written on the branches. The travel cost of the colored path is $z=c_{d2}+c_{24}+c_{43}+c_{31}+c_{1d}$. Then, this value is compared to the travel cost of the unexplored branches. If a branch has an initial travel cost lower than the current optimal value, the branch is explored next by the algorithm. \label{fig:TSPBB}}
\end{figure}
\endgroup

\section{The QUBO and Ising formulation}\label{sec:III}

In this work we analyze the use of the D-wave Advantage quantum computer to solve the combinatorial problems introduced in the previous sections. The D-wave machine belongs to the class of quantum annealers, that work through the application of a global time-dependent Hamiltonian. An introduction to quantum annealers is presented in Appendix A. 

The D-Wave machine evolves with an Ising-like Hamiltonian,  

\begin{eqnarray}\label{eq:IsingHamiltonian}
H(s)&=&-A(s)\sum_i\hat{\sigma}_x^{i}\\
&&+B(s)\bigg{(}\sum_ih_i\hat{\sigma}_z^{i}+\sum_{i>j}J_{ij}\hat{\sigma}_z^{i}\hat{\sigma}_z^{j}\bigg{)}.\nonumber
\end{eqnarray}

\noindent The coefficient $A(s), B(s)$ have the role of the schedule function $f$ in Eq.~\eqref{eq:adiabaticHamiltonian}, and $s$ is the adimensional time $s=t/t_a$, normalized with respect to the \textit{annealing time}. 

We can write any binary linear problem with constraints as an Ising problem.
In fact, any BLOP can be written as a quadratic unconstrained binary optimization (QUBO) problem. This defines a new cost function

\begin{equation}\label{eq:QUBO}
Q=-\sum_{i=1}^{N+n'}c_ix_i+\sum_{j=1}^m\lambda_j(b_j-\sum_{i=1}^{N+n'}A_{ij}x_i)^2,
\end{equation}

\noindent where we have introduced new $n'=\log_2(\max(b_j)+1)$ binary slack variables, that make the problem unconstrained. Each additional $i$-th slack variable, with $i=N+1,\dots,N+n'$ has cost coefficient $c_i=0$ and constraint matrix components $A_{ij}=2^{i-1}/2^N$, with $j=1,\dots,m$.
The $m$ parameters $\lambda_j$ are called Lagrange multipliers. Generally speaking, if $\lambda_j$ are too small, the minimization of the first term is favored, which corresponds to the minimization of the cost function, without any constraints. On the other hand, if $\lambda_j$ is too large, the second term acquires more importance and the optimal solution tends to satisfy the $j$-th constraint, ignoring the other terms. There is a range of values for each $\lambda_j$ s.t. the optimal solution of $Q$ is the optimal solution of Eq.\eqref{eq:BLproblem}. However, this range of values is both problem and size dependent.

In order to pass from binary variables into spin variables we define $M=N+n'$ new variables
\begin{eqnarray}\label{eq:binarytospin}
s_i& = &\frac{2x_i-1}{2},\quad s_i\in\{-\frac{1}{2},\frac{1}{2}\},
\end{eqnarray}
and set 
\begin{eqnarray}\label{eq:binarytospin_parameters}
h_i& = &\frac{c_i}{2}+\sum_{j=1}^m\lambda_j b_j A_{ij}-\sum_{j=1}^m\frac{\lambda_j}{2}A_{ij}\sum_{k=i}^{M}A_{kj}\nonumber\\
J_{ij}& = & \sum_{k=1}^m\frac{\lambda_k}{2}A_{ik}A_{jk},
\end{eqnarray}

\noindent that transform the QUBO function $Q$ of Eq.\eqref{eq:QUBO} into

\begin{equation}
H_Q=\sum_ih_is_i+\sum_{i>j}J_{ij}s_is_j,
\end{equation}

\noindent whose minimal energy solution corresponds to the ground state of the quantum Hamiltonian~\eqref{eq:IsingHamiltonian} at $t=t_a$.

Although we have a variety of results that show the power of quantun annealers, other results show that when the number of qubits is large the D-Wave machine has difficulty into finding the global solution~\cite{Pusey2020}. In the next section we are going to propose a way to circumvent this problem. 

\section{Efficient use of quantum annealers in hybrid classical-quantum algorithm}\label{sec:IV}

In this section we propose an hybrid way to use the currently available quantum annealers to produce reliable solution to some NP-hard problems. 
In Section~\ref{sec:II} we explained how the BB algorithm can be used to treat either the KP and the TSP. Here, we apply this algorithm to both the problems, stopping when the restricted sub-problems have a size that is small enough that the optimal solution can be obtained by the D-Wave machine.

\subsection{The knapsack problem}

Suppose we want to solve a KP with $N$ objects and capacity $W$. 
In order to be able to tract the optimal solution of problem~\eqref{eq:KP}, we choose $N$ objects with increasing value $v_i=i$, and with same weight $w_i=1$, for $i=1,\dots,N$. Thus, the optimal solution is the knapsack filled with just the last $W$ objects with highest value, $\mathbf{x}_{\text{opt}}=(0,\dots,0,\bold{e}_W)$, with $\bold{e}_W$ being the $W$-dimensional vector of ones, $\bold{e}_W=(1,\dots,1)$, and with the total value $z_{\text{opt}}=W[N+\frac{1}{2}(1-W)]$. After including the slack variables, the vector $\mathbf{x}_{\text{opt}}$ has $M=N+\lceil{\log_2(W+1)}\rceil$ components.

The QUBO function~\eqref{eq:QUBO} is

\begin{equation}
Q = - \sum_{i=1}^M v_ix_i+\lambda(W-\sum_{i=1}^Mw_ix_i)^2.
\end{equation}

Using Eqs.~\eqref{eq:KP} and \eqref{eq:binarytospin} we can write the Ising Hamiltonian of the problem. Because Eq.~\eqref{eq:KP} has only one constraint, we just need to adjust the one parameter $\lambda$. In order to find the optimal value of $\lambda$ we proceed as follows. 

We suppose $\mathbf{y}$ is a solution that satisfies the constraint. If we add a single object, say $x_l$ that overloads the knapsack, the QUBO function should be s.t. 
\begin{equation}
Q(\mathbf{y}+x_l)>Q(\mathbf{y}). \label{eq:lambda_condition}
\end{equation}

\noindent As it has to be valid for any possible solution, we need to take the maximum of the right side of Eq.~\eqref{eq:lambda_condition}. Thus, we set $\lambda=\max_i v_i+1$.
This choice ensures that the Hamiltonian ground state is also the optimal solution of the KP. 

\noindent We have performed an analysis of the bandgap for this particular configuration and we have seen that it scales polynomially with the size of the problem as $M^{-\alpha}$ for some exponent $\alpha$, see Appendix B. This is a promising feature, that means the total annealing time scales as $t_a\sim M^\alpha$ and therefore makes the problem solvable in polynomial time. We have to stress the fact that the polynomial scaling refers to this particular configuration of the problem, and it is not applicable to all the possible KPs.  

Let us first examine what happens when we solve the {\it unrestricted BLOP} on the D-Wave machine. In this case, we find that not only the solution is not the optimal one but also the constraint is not satisfied.  Indeed, when the number of qubits is too large, the system is not able to act as an ideal quantum annealer~\cite{Pusey2020}, as the results are affected by thermal noise~\cite{Amin2015}. One may think of increasing the value of the parameter $\lambda$ in order to force the system to prefer the fulfillment of the constraint. After several trials we have seen that this is not enough. Furthermore, the effect of thermal fluctuations leads the quantum annealers solution to be almost independent on the annealing time, reaching a plateau of the probability $p_0$ of measuring the ground state.
This is shown in Fig.~\ref{fig:p0KP}, where we plot $p_0$ obtained as the frequency of the ground state appearing as outcome, measured out of $1000$ reads for different annealing times $t_a$, calculated for a KP with $N$-objects and capacity $W=N$.

\begingroup
\begin{figure}
\centering
\includegraphics[width=.8\linewidth]{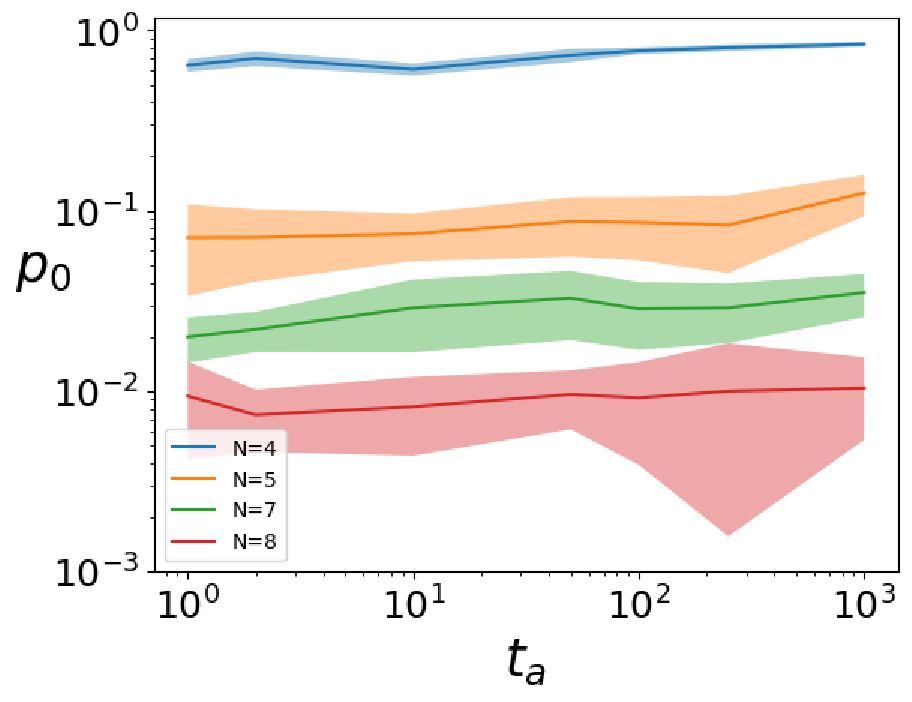}
\caption{Measured probability $p_0$ as a function of the annealing time for different $N$-objects knapsack problems with $W=N$. The value is an average over 20 runs, with 1000 measurements per run.\label{fig:p0KP}}
\end{figure} 
\endgroup

\begingroup
\begin{figure}
\centering
\includegraphics[width=.8\linewidth]{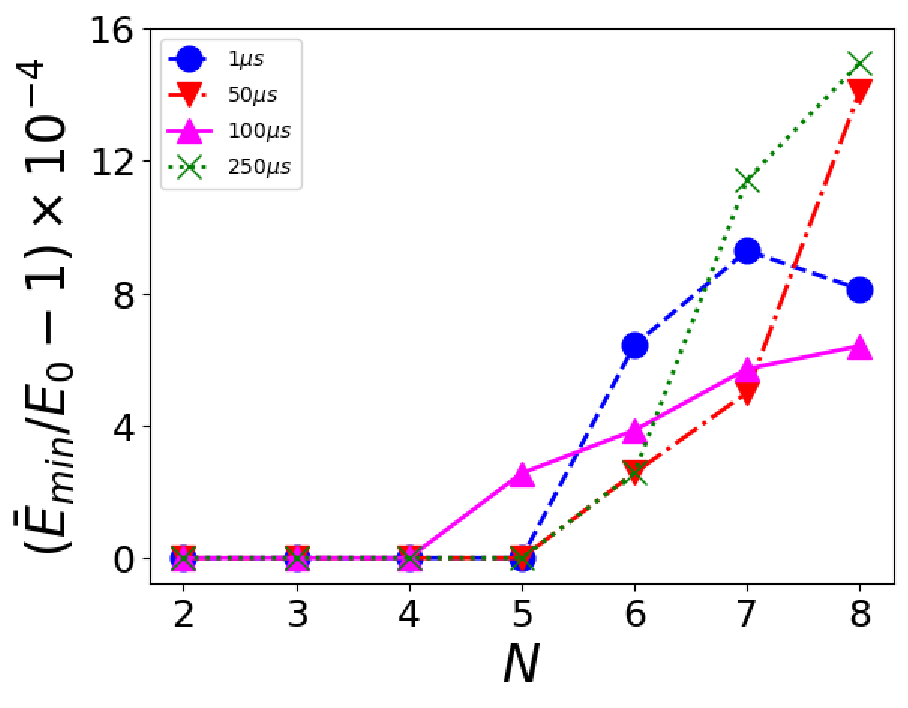}
\caption{Mean value of the minimal energy found for the $N$-objects knapsack problems with $W=N$ for different annealing times. The value is an average over 20 runs, with 1000 measurements per run.\label{fig:EminKP}}
\end{figure} 
\endgroup

\noindent In Fig.~\ref{fig:EminKP} we show also the mean value of the minimal energy found on 20 runs of the quantum annealer, each with 1000 measurements. We see the that the result is independent of the annealing time. 
A similar result was found in \cite{Krauss2020}. The authors analyzed another combinatorial problem, the shortest path problem, with the D-Wave quantum processor and they did not find correspondence between the annealing time and the frequency nor the energy distribution. This means that the behavior of the processor is not dependent on the annealing time, although the theory says otherwise. These results allow us to fix our annealing time to the value of $10\mu s$.

To better analyze the quality of the results obtained by the quantum annealer, we can introduce the following three figures of merit that estimate the quality of the obtained solution $\mathbf{x}_a$ with respect to the optimal one $\mathbf{x}_{\text{opt}}$:\\
- the normalized knapsack value distance 
\begin{equation}\label{eq:deltav}
\Delta \tilde{v}=\frac{z(\mathbf{x}_a)-z(\mathbf{x}_{\text{opt}})}{z(\mathbf{x}_{\text{opt}})},
\end{equation}
- the normalized knapsack weight 
\begin{equation}\label{eq:wtilde}
\tilde{w}=\frac{w(\mathbf{x}_a)}{W},
\end{equation}
- the Hamming distance $H$~\cite{NielsenChuang2000}
\begin{equation}\label{eq:Hamming}
H(\mathbf{x}_a,\mathbf{x}_{\text{opt}}) = \sum_{i=1}^M(\mathbf{x}_a^i\oplus\mathbf{x}_{\text{opt}}^i),
\end{equation}
where $M$ is the string length of the solutions and $\oplus$ stands for the sum modulo 2.

\begingroup
\begin{figure}
\centering
\includegraphics[width=.8\linewidth]{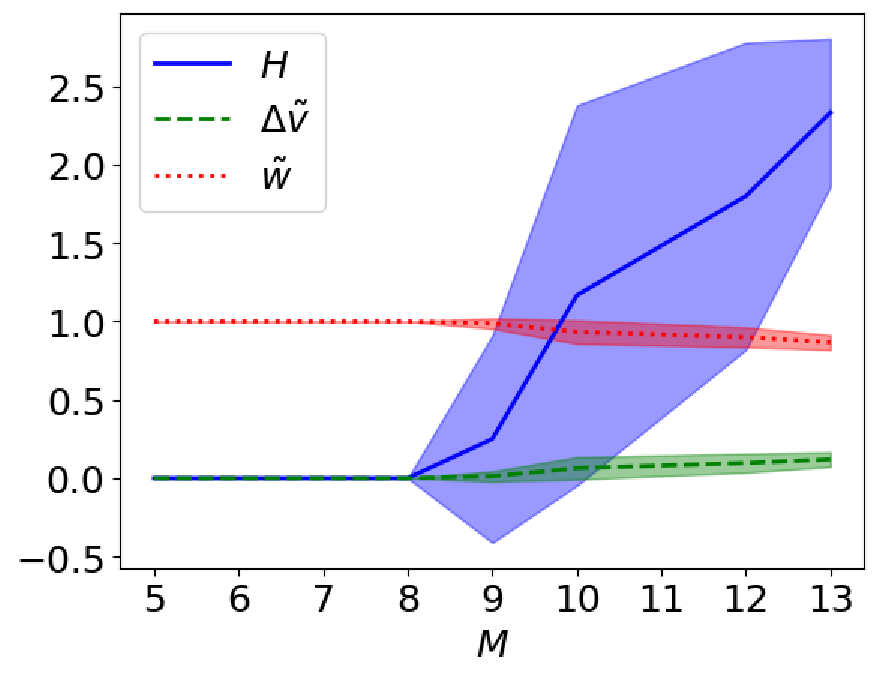}
\caption{The figure shows $\Delta\tilde{v}$, $\tilde{w}$, and $H$ of Eqs.~\eqref{eq:deltav},\eqref{eq:wtilde} and~\eqref{eq:Hamming} respectively, between the solution found by the quantum annealer $\mathbf{x}_a$ and the optimal solution $\mathbf{x}_{\text{opt}}$. The shaded regions accounts for the variances calculated over 20 runs. \label{fig:NeqW}}
\end{figure} 
\endgroup

Fig.~\ref{fig:NeqW} shows these metrics as a function of the number of qubits $M$ used by the quantum annealer. The width of the shaded region is the variance calculated over 20 different runs. We see that already for $M>8$ the result can differ from the optimal solution and have a variance different from zero. 

A good trade-off between the certainty of the result and the size of the problem can be provided by the {\it BB algorithm}, that divides the problem into sub-problems that are described by a lower number of qubits, for which the quantum annealer can provide more reliable solutions. 

The BB algorithm applied to the knapsack problem, as explained in section~\ref{sec:II}, explores a number of nodes that goes as $\mathcal{O}(2^N)$, since any time it explores a new branch, it creates a new KP with one less object and an updated knapsack capacity.  
In the toy problem we are considering, the number of nodes that saturate the constraint are $\begin{pmatrix}
N\\
W
\end{pmatrix}=\frac{N!}{W!(N-W)!}$, that is the number of combinations with $W$ objects chosen from a set of $N$. 
This corresponds to the exploration by the KP-BB algorithm of a number of branches
\begin{equation}
n_b = \sum_{k=1}^W\begin{pmatrix}
N\\
k
\end{pmatrix}.
\end{equation}

\noindent This value is lower than the exponential value obtained by applying the BB algorithm depicted in Fig.~\ref{fig:BB}, and the reason why is that the optimized KP-BB algorithm makes restrictions on more than one variable at once. But there is no point to make use of a quantum processor, since, for each node, all the variables have been set to a fixed value. 

Here, we propose an alternative scheme: we can devise a hybrid classical-quantum protocol by first exploiting the advantage of the BB algorithm to reduce the problem down to a chosen size $N'$, which corresponds to a number of variables $M=N'+\log_2(W'+1)$, given by the number of remaining objects $N'$ and by the remaining loading capacity of the knapsack $W'$. 
Then we solve the residual problem with the quantum annealer, which might be efficient to solve problems with a relatively small number of qubits.

We consider a KP with $N=25$ objects and loading capacity $W=10$. The problem can be described as QUBO by a number of binary variables $M_{\mathcal{P}}=29$. 

\begingroup
\begin{figure}
\centering
\includegraphics[width=.8\linewidth]{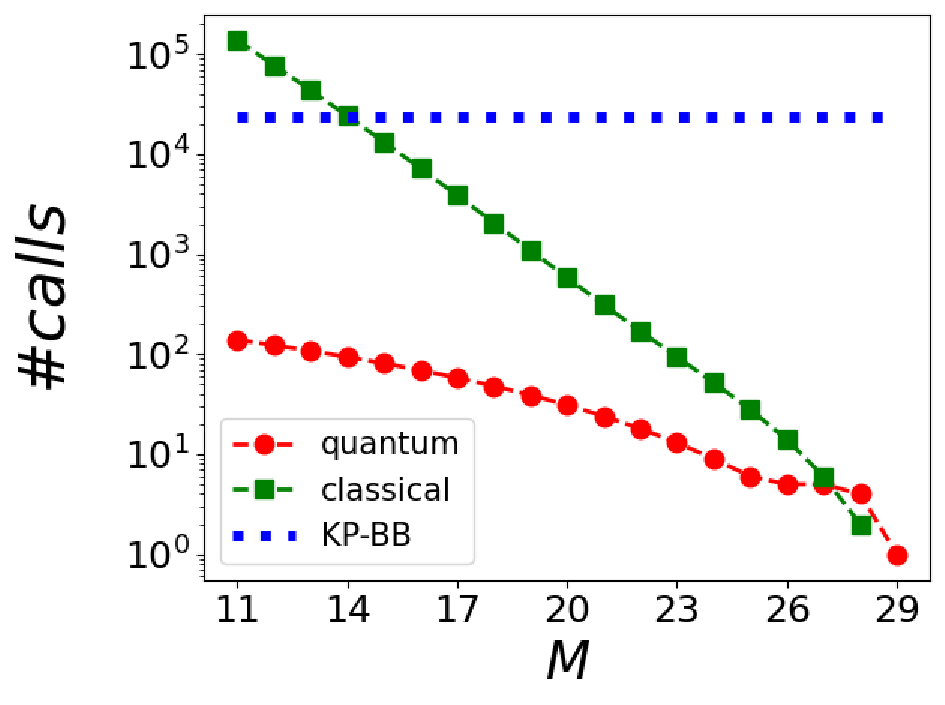}
\caption{The number of steps of the hybrid-BB algorithm performed on the classical (green square points) and quantum (red circle points) computers for a KP with $W=10$ and $N=25$ as a function of the maximum number of qubits $M$ used by the quantum annealer. The classical calls count the branching and the bounding procedures, while the quantum calls count the queries to the quantum annealer.
This has to be compared with the number of times the fully classical KP-BB algorithm performs branch--and--bound (dotted horizontal blue line). \label{fig:num_calls_KP}}
\end{figure} 
\endgroup

Fig.~\ref{fig:num_calls_KP} shows the number of branches explored by the BB algorithm (green squares) and the number of calls made to the quantum annealer (red dots) as a function of the chosen size $M$ The horizontal dotted blue line represents the number of branches explored by the optimized KP-BB algorithm, which is constant since it depends on $N, W$ only. We see that when the number of qubits $M>14$, the number of times the branch--and--bound procedure is applied in the hybrid algorithm is lower than the number of times this is done in the optimized KP-BB algorithm. Each reduced problem is eventually solved by the quantum annealer. In the extreme case of $M=M_{\mathcal{P}}$, the number of branches operated by the classical algorithm is one, as the original problem is defined and promptly sent to the quantum annealer.

Figs.~\ref{fig:N25W10v},~\ref{fig:N25W10w}, and~\ref{fig:N25W10H} show the metrics defined in Eq.ns (\ref{eq:deltav},\ref{eq:wtilde},\ref{eq:Hamming}) for the best obtained solution $\mathbf{x}_a$ as a function of the number of available qubits $M$, for a KP with $N=25$ and $W=10$. The width of the shaded region is the variance obtained over 20 different runs. The results are compared with a random outcome of strings, constructed as follows. In order to make a proper comparison, we extract randomly 1000 strings, as many as the number of measurements taken per sample. Then we choose as optimal solution the one that has a lower energy, in analogy with the procedure of the D-Wave machine. We then repeat the procedure 1000 times in order to get the statistics.  


\begingroup
\begin{figure}
\centering
\subfloat[]{\label{fig:N25W10v}\includegraphics[width=.8\linewidth]{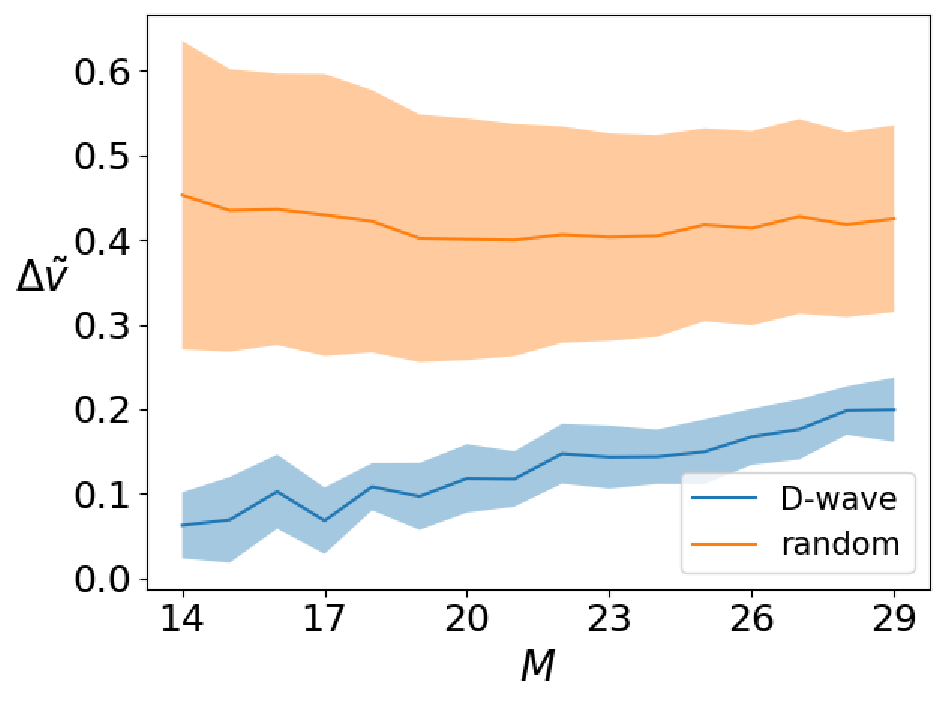}}

\subfloat[]{\label{fig:N25W10w}\includegraphics[width=.8\linewidth]{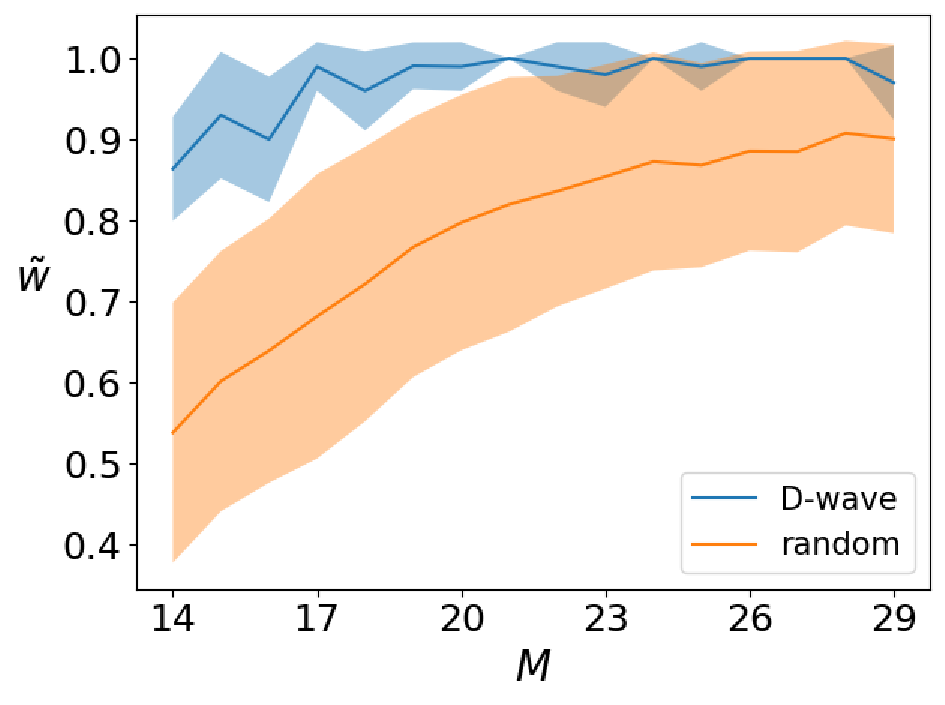}}

\subfloat[]{\label{fig:N25W10H}\includegraphics[width=.8\linewidth]{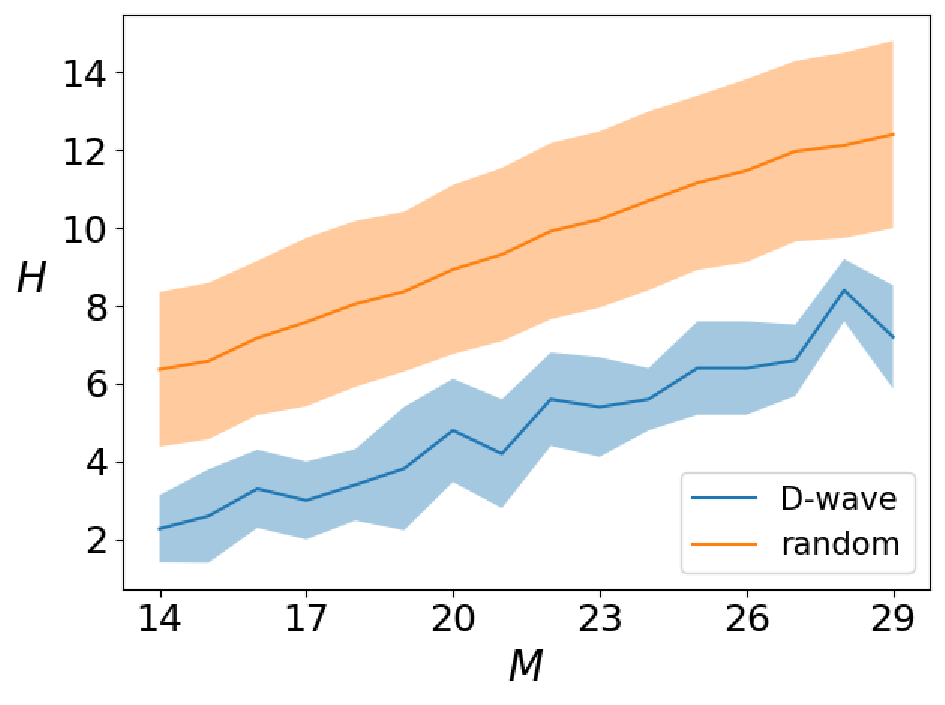}}
\caption{(a)The normalized value distance $\Delta\tilde{v}$, (b) the normalized knapsack weight $\tilde{w}$, and (c) the Hamming distance $H$, between the best solution obtained by the quantum annealer and the optimal solution obtained for a KP with $N=25$, $W=10$, solved with the hybrid BB algorithm with $M$ qubits and averaged over 20 runs. The result is compared to what is obtained by random instances as described in the text.\label{fig:N25W10}}
\end{figure} 
\endgroup

We see that the outcomes of the quantum annealer are not comparable with random guesses. Figs.~\ref{fig:N25W10v}  and~\ref{fig:N25W10w} show that the solution found by the D-wave machine has cost value closer to the optimal one and loading closer to the threshold set by the loading capacity. Finally, Fig.~\ref{fig:N25W10H} shows that the Hamming distance is lower than the one reached by random outcomes. 

Depending on what is our tolerance with respect to the different quality measures $\Delta\tilde{v},\tilde{w}$, and $H$, we may decide to restrict our problem till a certain number of qubits, thus speeding up the resolution of the problem. 
\begingroup
\begin{figure}
\subfloat[]{\includegraphics[width=.8\linewidth]{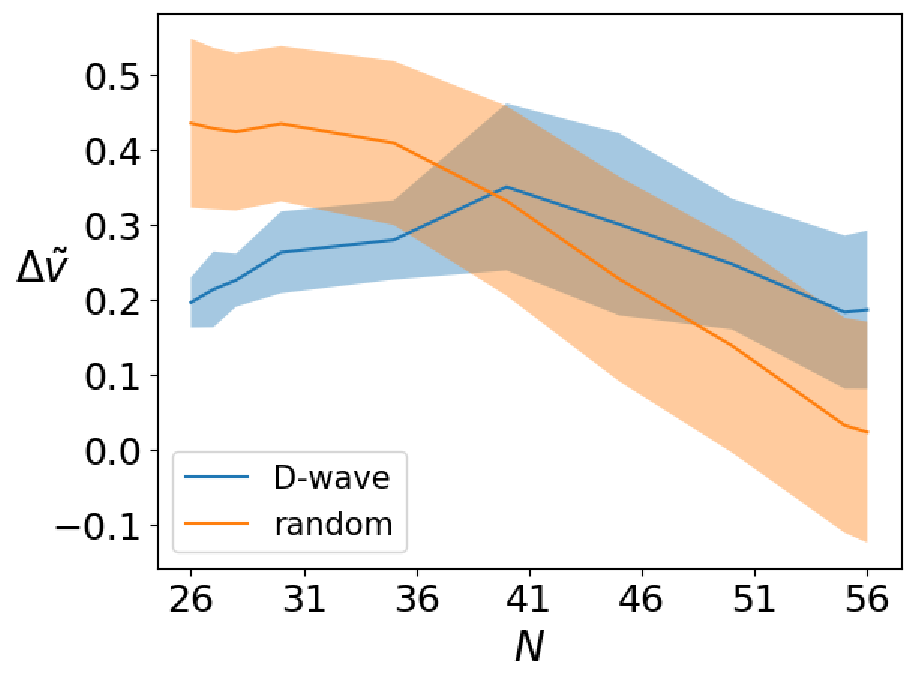}\label{fig:NW10v}}

\subfloat[]{\includegraphics[width=.8\linewidth]{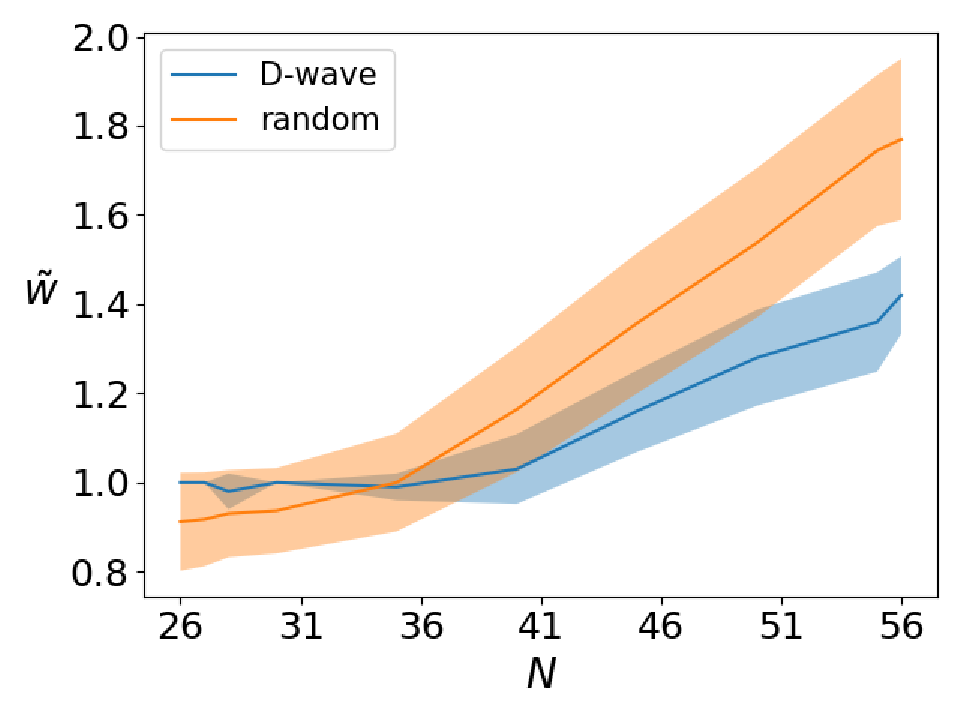}\label{fig:NW10w}}

\subfloat[]{\includegraphics[width=.8\linewidth]{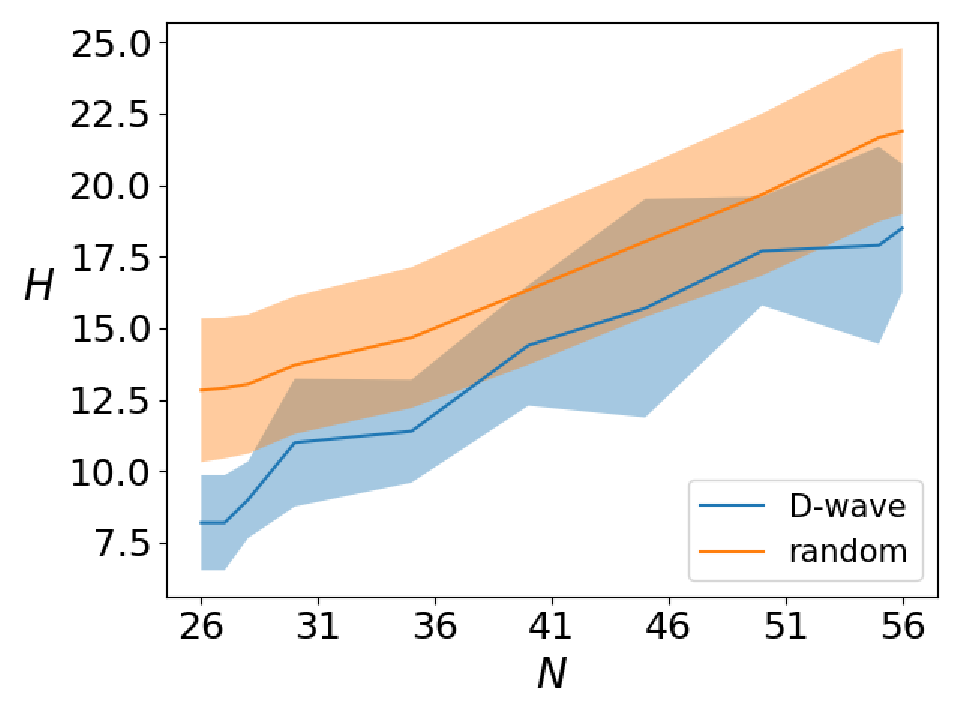}\label{fig:NW10H}}
\caption{(a)The normalized value distance $\Delta\tilde{v}$, (b) the normalized knapsack weight $\tilde{w}$, and (c) the Hamming distance $H$, between the best solution obtained by the quantum annealer and the optimal solution obtained for a KP with varying $N$ and fixed $W=10$, solved with the hybrid BB algorithm with $M=N+4$ qubits and averaged over 20 runs. The result is compared to what is obtained by random instances as described in the text.\label{fig:NW10}}
\end{figure} 
\endgroup

Figs.~\ref{fig:NW10v},~\ref{fig:NW10w}, and~\ref{fig:NW10H} show the same experiment performed keeping $W=10$ and varying the number of objects $N$, using the maximum number of necessary qubits $M=N+4$. 
Although the optimal solution is never reached, the D-wave machine performs better than random guessing even in this case, as it is evident from the results of the metrics. In fact, although the average $\Delta \tilde{v}$ of the D-wave outcomes is larger then the corresponding value for random guessing, Fig.~\ref{fig:NW10v}, this is mainly due to the fact that the machine tries to satisfy the capacity constraint, as shown in Fig.~\ref{fig:NW10w}. Hence, it prioritizes a lower number of objects over the desire for a higher load value. The better performance of the D-wave machine is captured by the lower value of the Hamming distance in Fig.~\ref{fig:NW10H}.

\subsection{The traveling salesman problem}

Let us move now to solve a TSP with $N$ cities using the quantum annealer, by  means of a hybrid classical-quantum protocol similar (but not equal) to the one presented for the KP.

The first step is to write the QUBO formulation of the problem as a function of a $N^2$ binary vector $\mathbf{x}$ with components $x_{i,j}$, with $i,j=1,\dots,N$. We use here the formulation suggested in Ref.~\cite{Lucas2014}, such that the first subscript indicates the city, and the second indicates the step. Thus, $x_{i,j}=1$ if the $i$-th city is visited at step $j$. Followig Eqs.~\eqref{eq:TSP} we can write the QUBO function \eqref{eq:QUBO} as~\cite{Lucas2014}

\begin{eqnarray}
Q^{\text{TSP}}(\mathbf{x})&=&\sum_{i,j=1}^NC_{ij}\sum_{k=1}^Nx_{i,k}x_{j,k+1}\nonumber\\
&&+\lambda\sum_{i=1}^N\bigg{(}1-\sum_{j=1}^Nx_{i,j}\bigg{)}^2\nonumber\\
&&+\lambda\sum_{j=1}^N\bigg{(}1-\sum_{i=1}^Nx_{i,j}\bigg{)}^2,\label{eq:QUBOTSP}
\end{eqnarray}

\noindent where the choice $\lambda>\max_{ij}C_{ij}$ ensures that the state with minimal energy is the optimal solution of the TSP.

The TSP-BB algorithm differs from the KP-BB algorithm as it has to cross all the $N$ cities. The maximum number of branches explored by the TSP-BB algorithm is $\sum_{n=1}^{N-1}(N-n)!$. If we decide to stop the BB algorithm when the total size of the problem is $M^2$ (with $M$ cities left), the number of branches explored by the BB algorithm would be at maximum $\sum_{n=M}^{N-1}(N-n)!$, and the total number of calls to the D-Wave machine will be in the worst case scenario $N!/M!$.
However, for the large majority of the problems we have examined, the D-Wave processor is called a very small number of times.

Let's consider a scenario of $N=10$ cities, all connected together by a route, with a non symmetric cost matrix given by $C_{ij}=(i-j)\text{mod}N$. The optimal solution $\mathbf{x}_{\text{opt}}$ of this problem has components $x_{ij}=1$ if $i=j$, and $0$ otherwise. This means that the nodes are crossed in the same order that they are labeled.
 
We have applied the hybrid algorithm to solve this problem and we have obtained a number of calls of the quantum hardware equal to $10$ for a number of qubits between $36$ ($M=6$) and $81$ ($M=9$).
Clearly these numbers are problem-dependent, but they give a good indication of the small number of calls to the quantum hardware when the hybrid approach is used. 

For this problem, as figure of merits of the quality of the protocol we now use:\\
- the travel cost of the found solution $\mathbf{x}_a$, normalized to the travel cost to the optimal solution $\mathbf{x}_{\text{opt}}$:
\begin{equation}
\tilde{c}=\frac{z(\mathbf{x}_a)}{z(\mathbf{x}_{\text{opt}})}
\end{equation} 
- and the Hamming distance $H=H(\mathbf{x}_a,\mathbf{x}_{\text{opt}})$ of  Eq.~\eqref{eq:Hamming}.\\
These two metrics are shown in Fig.~\ref{fig:N10TSPc} and Fig.~\ref{fig:N10TSPH} respectively, as function of the chosen number of qubits $M$. As in the previous section, the results are compared with a random path throughout the cities. Contrary to the KP problem, the random solutions can be chosen here to strictly satisfy the constraints, since it suffices to generate the solution with a random ordering of the $M$ cities.

\begingroup
\begin{figure}
\subfloat[]{\includegraphics[width=.8\linewidth]{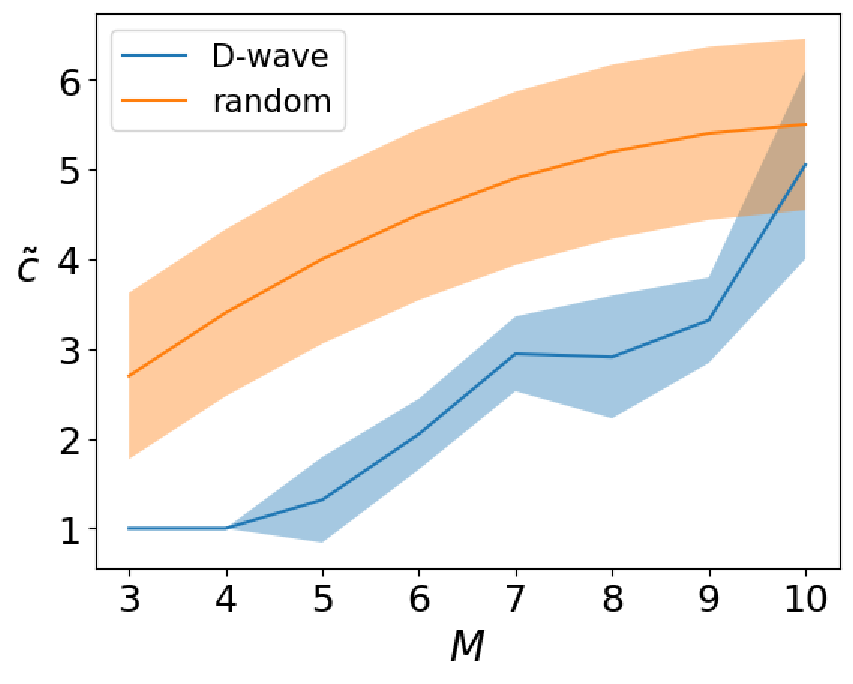}\label{fig:N10TSPc}}

\subfloat[]{\includegraphics[width=.8\linewidth]{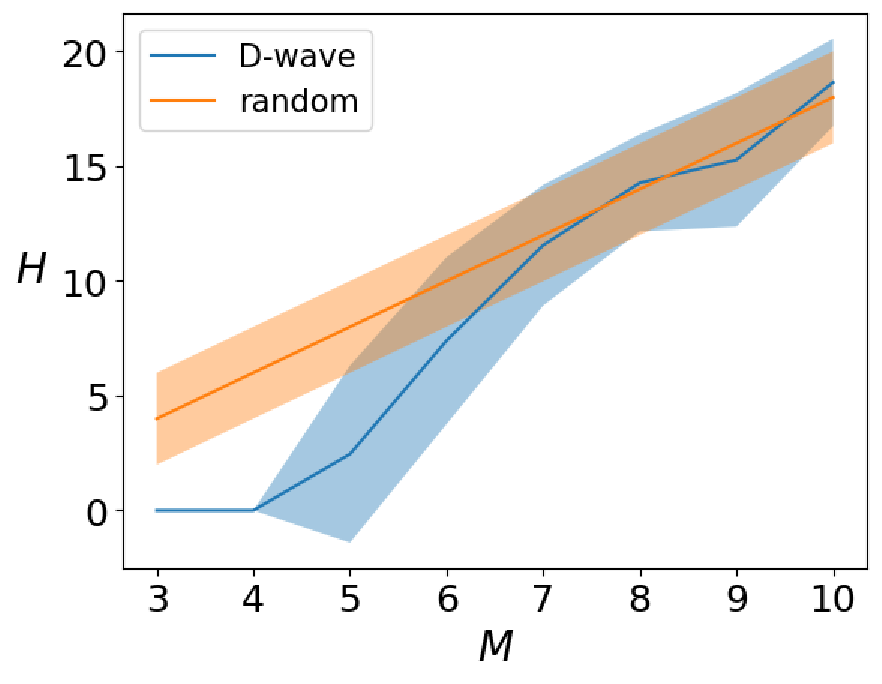}\label{fig:N10TSPH}}
\caption{(a)The normalized travel cost $\tilde{c}$, (b) the Hamming distance $H$, between the best solution obtained by the quantum annealer and the optimal solution, obtained for a TSP with $N$  cities solved with the hybrid BB algorithm with $M^2$ qubits and averaged over 20 runs. The result is compared to what is obtained by random instances as described in the text.\label{fig:N10TSP}}
\end{figure} 
\endgroup

The normalized travel cost is lower than the one obtained for random paths for all the instances analyzed, up to a maximum of 100 qubits ($M=10$). We see that the optimal solution has been obtained only for the cases with $M=3,4$, whereas for larger TSP the D-wave machine was not able to find it. Also, comparing Figs.~\ref{fig:N10TSPc} and~\ref{fig:N10TSPH}, we see that the Hamming distance alone does not effectively measure the quality of the solution. Already for a TSP with $M=7$ cities, the results are not distinguishable from random outcomes in terms of $H$, yet they exhibit a significantly lower travel cost value.

\section{Conclusion}\label{sec:V}

In this paper we have investigated the resolution of two binary linear problems, the knapsack problem and the traveling salesman problem, that are known to be NP-hard. Those are often solved with the branch--and--bound algorithm that we have described in Sec.~\ref{sec:II}. 
After introducing the two problems we have moved our attention to the quantum annealer in Sec.\ref{sec:III}, that is a quantum computer that offers a global method for resolution of binary linear problems.
In this work we have merged the classical and quantum method for resolution of those NP-hard problems, in order to show the advantage of a hybrid approach. In fact, branch--and--bounds algorithm do not offer a significant speed up with respect to brute force strategies, whereas quantum annealers suffer from a low reliability when the number of instances is large. 
However, when merged together, branch--and--bound defines a new problem with a smaller size that can be handled by the global quantum solver. 
Although there is a long way to go with quantum annealers, we believe this work shows a good strategy to use them.

{\appendix

\section*{Appendix A\\Quantum Annealers}\label{app:annealers}
Quantum annealers are a class of quantum computers which make use of a time-dependent evolution Hamiltonian to solve combinatorial problems. 

The adiabatic theorem~\cite{Kato1950,Albash2018} is the basis of the computation with quantum annealers. Suppose that a quantum system is subject to a time evolving Hamiltonian $H(t)$, and at time $t$, the system is in the $j$-th eigenstate $|\epsilon_j(t)\rangle$, such that $H(t)|\epsilon_j(t)\rangle=\epsilon_j|\epsilon_j(t)\rangle$. 
The eigenvalues $\epsilon_j$ are sorted in increasing order and non-degenerate. 
The adiabatic theorem states the if the evolution is performed \textit{slowly enough}, the quantum system stays in the evolved $j$-th instantaneous eigenstate of the Hamiltonian $H(t)$. 
Understanding what "slowly enough'' means is the role of the adiabatic theorem. Several versions of the theorem have been developed during the years, depending on the initial conditions of the system. 
The Kato version~\cite{Kato1950} of the adiabatic theorem gives a lower bound for the annealing time $t_a$ in terms of the first and second derivatives of the Hamiltonian $H(t)$, and in terms of the minimum value assumed during the evolution by the energy gap $\Delta_{ij}=|\epsilon_j-\epsilon_i|$, $\forall i\neq j$. 
An approximate version of the theorem yields the inequality~\cite{Albash2018}

\begin{equation}\label{eq:adiabatic_theorem}
\max_{s\in[0,1]}\frac{|\langle\epsilon_i(s)|\partial_sH(s)|\epsilon_j(s)\rangle|}{\Delta_{ij}(s)^2}\leq t_a,\quad \forall i\neq j.
\end{equation}

Since we are interested in the ground state, for our purposes, we consider only the energy gap between the instantaneous ground state and the first energy level $\Delta=|\epsilon_1-\epsilon_0|$.  

The time-dependent Hamiltonian $H(t)$ is composed by two terms, namely
\begin{equation}\label{eq:adiabaticHamiltonian}
H(t) = [1-f(t/t_a)]H_0+f(t/t_a)H_1,
\end{equation}

\noindent where $H_0$ and $H_1$ are two non-commuting Hamiltonians and $t_a$ is the total annealing time. The evolution function $f$, also called schedule, is such that $f(s):[0,1]\to[0,1]$, $f(0)=0$ and $f(1)=1$.

The Hamiltonian $H_0$ has a ground state that is easy to prepare, whereas $H_1$ is a Hamiltonian for which the ground state corresponds to the classical optimal solution of the combinatorial problem. Thus $H_1$ is the term that encodes our optimization problem.

At time $t=0$ the system is prepared in the ground state of the Hamiltonian $H_0$. 
For a sufficiently long $t=t_a$, the adiabatic theorem ensures that the state of the system is the solution of our classical problem, that we will recover measuring the system at this time.

The energy gap plays a crucial role in adiabatic quantum computation, and understanding how it scales with problem size is essential. The efficiency of a quantum annealer compared to a classical algorithm largely hinges on the scaling of the band gap.
However, it is important to note that the schedule influences the derivative of the Hamiltonian and, consequently, impacts the satisfaction of the inequality~\eqref{eq:adiabatic_theorem}. A modification to the schedule could either enhance or diminish the quantum annealer's performance. In this study, we focus on the linear schedule $f(t/t_a)=t/t_a$ deferring optimal schedule analysis to future research.

In order to understand the computational advantage of the quantum annealer we need to define the \textit{quantum speed up}. One definition~\cite{Albash2018} takes into account the speed up with respect to existing algorithms. Another useful definition is the \textit{limited} quantum speed up~\cite{Albash2018}, which compares a quantum algorithm with its classical analogue, that is an algorithm that proceeds through the same steps in the classical regimes. 
The classical version of the adiabatic quantum computation is the simulated annealing where the thermal fluctuations allow the exploration of different solutions. 
Both the classical and quantum annealing algorithms usually find approximations to the global solutions, although we have evidence that quantum annealing reaches the global solutions more often~\cite{Kadowaki1998} and with shorter annealing time~\cite{Suzuki2009,Santoro2008,Heim2015}.

\section*{Appendix B\\The energy gap}

In this appendix we consider the Knapsack problem with fixed loading capacity $W$ and increasing number of objects $N\geq W$. As in we did in the main text, all the objects have weight 1 and they are sorted by increasing value $v_i=i$, for $i=1,\dots,N$. The number of qubits needed to describe the problem is $M=N+\lceil\log_2(W+1)\rceil$. 
\begin{figure}
\includegraphics[width=.8\linewidth]{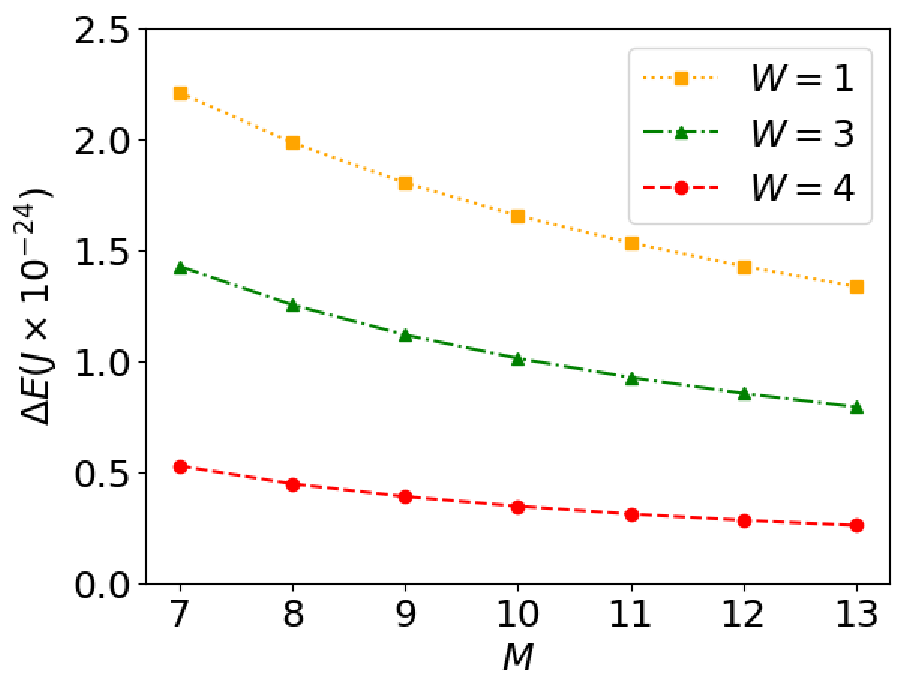}
\caption{The minimum of the ground state energy band gap $\Delta E$ varying the number of qubits $M$ for some KPs with fixed loading capacity $W$. \label{fig:DeltaE}}
\end{figure}
In order to be general, we have calculated the minimum value of the energy band gap $\Delta$, considering a one to one embedding into the hardware and considering the schedule functions $A,B$ of Eq.\eqref{eq:adiabaticHamiltonian} and their magnitudes as provided by D-Wave~\cite{Harris2010}. Fig.~\ref{fig:DeltaE} shows the minimum of the band gap that is reached during the annealing process. From our simulations, we see that the minimum of the band gap goes as a polynomial of the number of qubits $\Delta E\propto M^{-c}$. The polynomial scaling of $\Delta E$ is promising, as it reflects the polynomial scaling of the annealing time needed to find the optimal solution. 

However, this result is affected by the topology of the D-Wave quantum annealer.  In fact, the results that we found are applicable to an ideal completely connected quantum computer, where Hamiltonian \eqref{eq:adiabaticHamiltonian} can be directly implemented. Because of the constrained topology~\cite{Harris2010} of the machine, several ancillary qubits are used to embed the problem into the quantum annealer, therefore changing the scaling of the band gap. In this work we have used the embedding tool that is implemented in the SDK D-Wave Ocean. Hence we averaged the results over 20 runs, each with a different embedding onto the machine. 

\noindent Because of the embedding, We can not use the results that we have shown in this appendix to obtain the minimum value of the annealing time through the adiabatic theorem. However, we believe they provide an understanding of the problem and an hardware-agnostic analysis of the energy band gap.
}
\section*{Acknowledgment}

This research is funded by the International Foundation Big Data and Artificial Intelligence for Human Development (IFAB) through the project “Quantum Computing for Applications”.
E.~E.~is partially supported by INFN through the project “QUANTUM”. C.~S. and E.~E. acknowledge financial support from the National Centre for HPC, Big Data and Quantum Computing (Spoke 10, CN00000013).


\EOD
\end{document}